\documentclass[article,amsmath,amssymb]{revtex4}

\usepackage{graphicx}% Include figure files
\usepackage{dcolumn}% Align table columns on decimal point
\usepackage{bm}% bold math
\usepackage{amsmath,amssymb}
\begin{document}

%\preprint{APS/123-QED}

\title{FITTING POTENTIAL ENERGY SURFACE OF REACTIVE SYSTEM VIA GENETIC
ALGORITHM}
\author{Wiliam Ferreira da Cunha}\author{Luiz Fernando Roncaratti}
\altaffiliation{roncaratti@fis.unb.br}
\author{Ricardo Gargano}\author{Geraldo Magela e Silva}
\affiliation{Institute of Physics, University of Brasilia}

\date{\today}

\begin{abstract}

In this work, we present a new fitting of the $Na+HF$ potential energy surface (PES) utilizing a new optmization
method based in Genetic Algorithm. Topology studies, such as isoenergetic contours and Minimum Energy Path
(MEP), show that the quality of this new PES is comparable to the best PES of literature. These facts, suggests
that this new approach can be utilized as new tool to fit PES of reactive systems.

\end{abstract}

\maketitle

\section{Introduction}

%Through the last years, several types of numerical and combinatorial optimization algorithms have been used as
%useful tools to minimize functional forms. Generally, when those forms are non-linear or occur in problems
%without a specific optimization method, stochastic methods based on search algorithms have shown good results
%due to its smaller susceptibility to be trapped in a local minimum. Besides that, they can easily be implemented
%to work with other techniques. In this class of algorithms, the genetic ones have received special attention
%because they are a robust optimization tool. An algorithm can be named genetic when it uses some kind of
%codification to transform a set of possible solutions of a given problem in a population that will evolve
%subject to operators inspired, or not, by mechanisms of natural selection. In other words, they work with a
%population of solutions to obtain better solutions in the next generation. To do this, they use only information
%of cost and prize.

Genetic algorithms \cite{genetico1,genetico2,genetico3, genetico4, genetico5, genetico6} have been applied
successfully in the description of a variety of global minimization problems. It have as well attracted
significant attention due to their suitability for large-scale optimization problems, specially for those in
which a desired global minimum is hidden among many local minima.

The main object of this paper is to propose a genetic algorithm optimization technique(GAOT) for fitting
the PES through electronic energies obtained by {\it ab initio} calculations. In order to present and to
test the method, we reproduce the PES of the reaction involving HF and an alkali metal, namely
\begin{equation}
Na(3 {^{2}S_{1/2}}) + HF(X^{1}\Sigma^{+}) \rightarrow NaF(X^{1}\Sigma^{+}) + H(^{2}S_{1/2}), \label{sodium}
\end{equation}
utilizing as a trial function a Bond Order (BO) polynomial expansion \cite{bo1} as well as {\em ab initio}
calculations published by Lagan\`a {\em et al}\cite{bo2}. The motivation behind the choice of this reaction
is its high endoergicity and its bent transition state \cite{gargano1}. Further more, this reactive process has been
experimentally \cite{tech1,tech4}, theoretically \cite{gargano2,gargano4} and computational investigated \cite{gargano5}.

This paper is organized as follows. In Section 2, we present the main characteristics of the GAOT.
The details of the GAOT fitting and its comparison with other SEPs are shown in the Section 3.
Our conclusions are contained in the Section 4.

\section{Model}

\subsection{The Problem}

In order to fit a given functional form $V([\bold{a}],\vec{r})$ in some set of $n_p$ points $(\vec{r_p},e_p)$,
we want that the GAOT finds a set of parameters $[\bold{a}]=[a_1,a_2,..,a_m]$ that minimize the mean square
deviation
\begin{eqnarray}
S=\sum_{p}^{n_p}\delta_p^2=\sum_{p}^{n_p}(e_{p}-\overline{e}_{p})^2\label{rms}
\end{eqnarray}
where $\overline{e}_{p} \equiv V([\bold{a}],\vec{r_p})$.

\subsection{Codification}

In our genetic algorithm the population is coded in a binary discrete cube named $\bold{A}$, with $l\times
m\times n$ bits. The elements of $\bold{A}$, $a_{ijk}$, are either 0 or 1, with $i,j,k$ integers numbers $1\leq
i\leq n$, $1\leq j \leq m$, $1\leq k \leq n$. The label $i$ refers to the component $i$ of the {\it gene} $j$ of
the individual $k$. Therefore, $\bold{A}$ represents a population of $n$ individuals, each one of them have a
genetic code with $m$ {\it genes}. Each {\it gene} is a binary string with $l$ bits.

The genetic code of the individual $k$ is given by
\begin{equation}
[\overline{\bold{a}}]_k=[\overline{a}_{1k},\overline{a}_{2k},...,\overline{a}_{mk}],\nonumber
\end{equation}
were
\begin{equation}
\overline{a}_{jk}=\sum_{i=1}^{l}2^{i-1}a_{ijk} \label{aijk}
\end{equation}
is a integer number composed with the binary string $a_{1jk}a_{2jk}..a_{ijk}..a_{ljk}$. It is defined on the
interval $[0,2^{l}-1]$. To define the real search space for each parameter, we transform
\begin{equation}
\overline{a}_{jk} \rightarrow {a}_{jk} \equiv
\frac{(a^{max}_{j}-a^{min}_{j})}{2^{l}-1}\overline{a}_{jk}+a^{min}_{j}\label{ajk}
\end{equation}
were ${a}_{jk}$ is a real number defined on the interval $\delta_j=[a^{min}_{j},a^{max}_{j}]$.
\begin{equation}
\delta_j=[a^{min}_{j},a^{max}_{j}]\label{deltaj}.
\end{equation}

Now we define the phenotype of the individual $k$, $V_{k}\equiv V([\bold{a}]_{k},\vec{r})$ where
\begin{equation}
[\bold{a}]_{k}=[a_{1k},a_{2k},...,a_{jk},...,a_{mk}]\label{[ajk]}
\end{equation}
is a set of coefficients that characterize the individual $k$. With this we define the fitness of a phenotype
$k$
\begin{equation}
F_k=S_{max}-S_k\nonumber
\end{equation}
where
\begin{equation}
S_k=\sum_{p}^{n_p}(\delta_{kp})^2=\sum_{p}^{n_p}(e_{p}-V_{kp})^2\label{fitness}
\end{equation}
and $S_{max}$ is worst individual in the population. $V_{kp}\equiv V([\bold{a}]_{k},\vec{r_p})$ and
$\delta_{kp}$ is the difference among the {\it ab initio} energy $e_p$ and the fit of the individual $k$ in the
configuration $\vec{r}=\vec{r_p}$.

\subsection{Operators}

We use the most common operators: selection, recombination and mutation. The selection operator normalize the
vector $S_{k}$
\begin{equation}
P_{k}={\frac {S_{k}}{\sum S_{k}}}\label{prob}
\end{equation}
that represents the probability of each individual been selected for a recombination through a roulette
spinning. For the purpose of this work we selected $n/2$ individuals (parents) that will generate, through the
recombination operator, $n/2$ new individuals (offsprings). So, to make a new generation we joint the $n/2$ old
strings (parents) with a $n/2$ new strings (offsprings) in order to maintain a population with fixed number $n$.
The recombination operator is a cross-over operator that recombine the binary string of each gene $j$ of two
random selected individuals to form two new individuals. In this work we use a two random point cross-over.

The mutation operator flip $N_{mut}$ random selected bits in a population. We choose $N_{mut}$ to make the
probability of change of a given bit equal to $0.01$ per cent. So, in a population of $l\times m\times n$ bits,
we make
\begin{equation}
 q = \frac {N_{mut}}{l\times m\times n}\label{q}
\end{equation}
where $q$ is the probability of change in one bit.

The elitist strategy consists of copying an arbitrary number $N_{el}$ of the best individual on the population
in the next generation. It warrants that this individual will not be extinguished.

\section{Fitting the {\em ab initio} PES}

Lagan\`a {\em et al.} construed the BO5 PES of the Na+HF reaction considering a total of the 425 {\em ab initio}
values\cite{bo2}, being 42 values (Table 2 of the Ref.\cite{bo2}) calculated in the region that better
characterize the collinear Na-HF and F-NaH geometries (insertion). All these 425 energy values cover a relevant
portion of the surface at $\theta$, the angle formed by the NaF and HF internuclear distances, equal to
$\theta$=0$^{\circ}$, 45$^{\circ}$, 60$^{\circ}$, 75$^{\circ}$, 90$^{\circ}$, 120$^{\circ}$ and 180$^{\circ}$.
Analytical representations of the BO5 PES were obtained using a BO polynomial expansion for both two- and
three-body terms\cite{bo1}, following the standard many-body form. Each two-body term was construed fitting a
polynomial of the fourth order in the related BO variables. The three-body term was fitted using the BO
polynomial expansion given by
\begin{eqnarray}
V(R_{NaF},R_{HF},R_{NaH})=\sum_{x=0}^5\sum_{y=0}^5\sum_{z=0}^5a_{xyz}\eta _{NaF}^{x}\eta_{HF}^{y}\eta_{NaH
}^{z}\label{func}
\end{eqnarray}
with $x+y+z$ $\le 6$ and at least two indices differing from zero. The quantities $\eta_{m}$ are defined as
\begin{eqnarray}
\eta_{M} = e^{-\beta_{M}(R_{M}-R_{eM})}\label{eta}
\end{eqnarray}
with M=NaF, HF and NaH, where the parameters values of the Na+HF reactive process are
$R_{eNaF}=1.92595$ \AA, $R_{eHF}=0.91681$ \AA, $R_{eNaH}=1.88740$ \AA, $\beta_{NaF}=0.88260$ \AA$^{-1}$,
$\beta_{HF}=2.19406$ \AA$^{-1}$ and $\beta_{NaH}=1.19798$ \AA$^{-1}$.

In order to use the GAOT method to reproduce the Na+HF PES, we used only
the BO polynomial expansion for three-body term given by Eq. (\ref{func}), with the same powers utilized to fit
the BO5 PES. It should be pointed that we used only 243 of a total of the 425 {\em ab initio} values used to
produce the BO5 PES. This number was the same used in the GSA PES fitting \cite{gargano5} (see table table 1 of the
Ref.\cite{bo2}).

Using the notation of Section II and Eq. (\ref{func}) we define the phenotype of the individual $k$

\begin{eqnarray}
V_k\equiv V([\bold{a}]_k,R_{NaF},R_{HF},R_{NaH})=\sum_{j=1}^ma_{jk}\eta _{NaF}^{x_j}\eta_{HF}^{y_j}\
eta_{NaH}^{z_j}
\end{eqnarray}

where $m$ is the number of coefficients that we use in the expansion. Each coefficient $a_{jk}$ had a specific
fixed combination of powers $(x_j,y_j,z_j)$. In this way, we guarantee that the fittest individual in the
population had a set of coefficients $[\bold{a}]_{k}$ (\ref{[ajk]}) that better fit the Eq. (\ref{func}).

In this work, we define as an acceptable solution the set of 77 coefficients (\ref{ajk}) that fit the
expansion (\ref{func}) to 243 {\it ab initio} points in a way that the root mean square deviation (\ref{rms}) be
less than 1,0 Kcal/mol. In fact, we can find a large number of acceptable solutions. The set of all solutions is
the definition of search space ($\Gamma$). The length of $\Gamma$ is defined  by the number $m$ of coefficients
and the length $l$ of the binary codification. Each one of the 77 coefficients, that define the individual $k$,
can assume $2^l$ distinct values. So, an individual in the population is only one possibility among
$2^{l\times m}$. This value defines the length of $\Gamma$. The precision of $\Gamma$ describes the number of
digits that are used to express a real value $a_{jk}$ and shows the minimal difference between two
possible values of $a_{jk}$. Being each coefficient defined on an arbitrary interval $\delta_j$ (\ref{deltaj}),
the precision of the coefficient $a_{jk}$ is

\begin{eqnarray}
\frac{a_{j}^{max}-a_{j}^{min}}{2^l}.\label{precision}
\end{eqnarray}

If we do not have any information about the order of magnitude of the $a_{jk}$ values, we must choose the
intervals $\delta_j$'s such that they cover the greatest number of values. However, after some generations, we
obtained more precise information about the order of magnitude of each coefficient $a_{jk}$. In order to improve
the performance of the standard GAOT, we include in our technique the concept of dynamic search space. It
consist in the use of information of past generations to determine the length and precision of the search space
for the next generations. For the first generations, when we have few information about the order of magnitude
of the coefficients, we do not need many digits to represent a real number $a_{jk}$, that is, we use a low
precision codification given by a low value of $l$. In this way, we make $\Gamma$ a "small" search space and the
GAOT can find the regions of acceptable solutions faster. Once found some of these regions we can redefine the
intervals $\delta_j$'s and rise the precision rising the length of binary codification $l$. After extensive
trials of the parameters values we take $m=77$, $n=100$, $q=0,01$ and $N_{el}=10$. Beside that, we always start
the GAOT with a random population defined in the initial intervals
$\delta_j=[a_{j}^{min},a_{j}^{max}]=[-10^5,10^5]$ and set the initial value for the length of the binary
codification $l=12$. In this way we had a search space of length $2^{l\times m}=2^{12\times 77}=2^{924}$ and the
minimal difference of two possible values of $a_{jk}$ is $\frac{10^5}{2^{11}}\approxeq 49$. After 1000
generations we redefine $l=l+4$ and $\delta_j=[a_{j}^{min},a_{j}^{max}]$ where
$a_{j}^{min}=a_{jbest}+a_{j}^{min}\times 10^{-1}$, $a_{j}^{max}=a_{jbest}+a_{j}^{max}\times 10^{-1}$ and
$a_{jbest}$ is the fitest individual in the population found along the last 1000 generations. We set 10000
generations for each run of the GAOT. It should be pointed out that the algorithm is very robust and works
properly with an wide range of these parameters.

\begin{center}
\begin{figure}[h]
\includegraphics{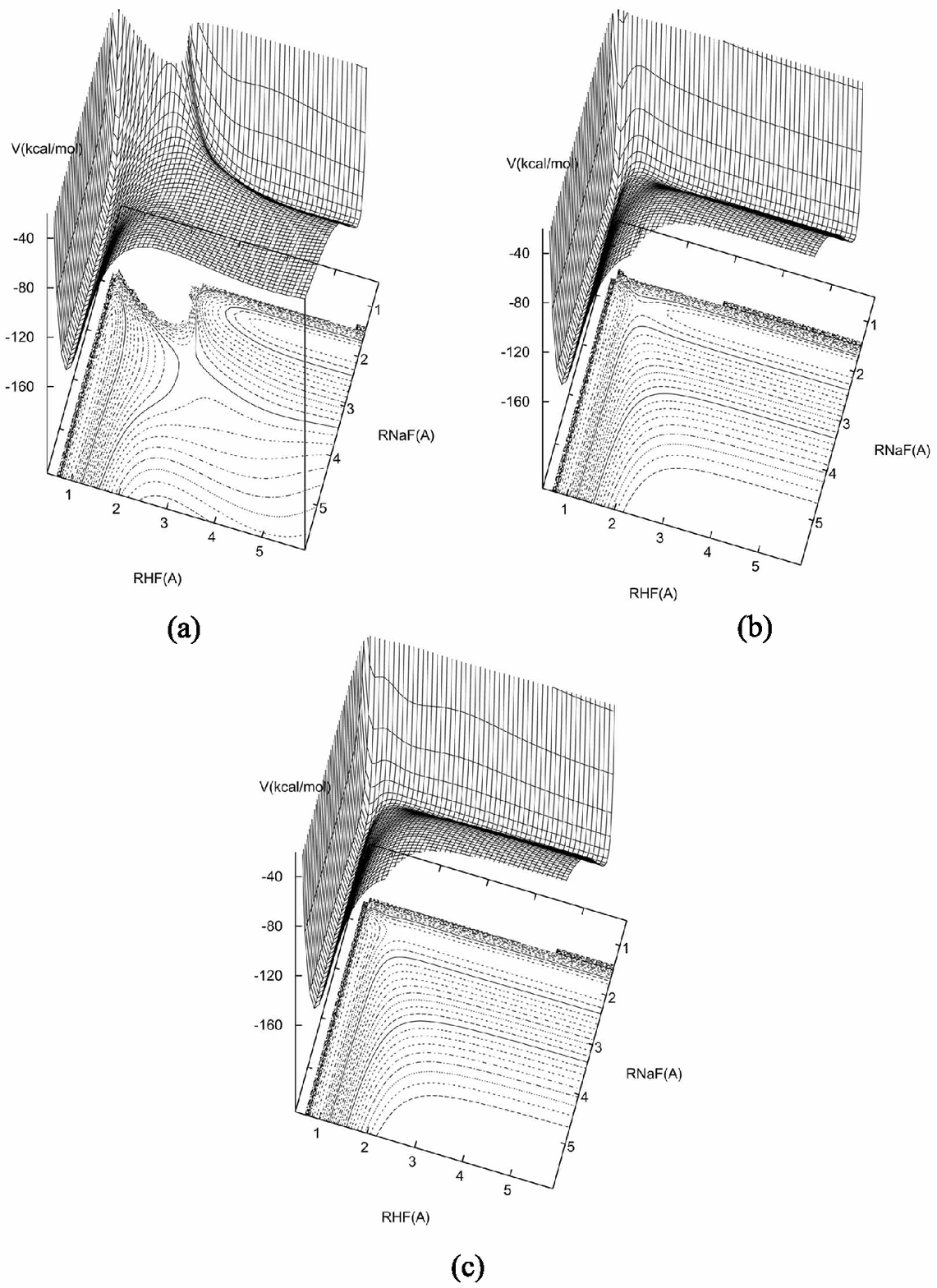}
\caption{The figure shows the isoenergetic contours of the GAOT PES, at $\theta$=30.0$^{\circ}$(a),
$\theta$=180.0$^{\circ}$(b) and $\theta$=77.2$^{\circ}$(c). The zero energy was set at the Na+HF asymptote. The
dashed contours are taken between -160 and -40 kcal/mol and them are spaced each other by 5 kcal/mol.}
\label{sep1}
\end{figure}
\end{center}

Coefficients and powers of the polynomial given by Eq.\ref{func} for the GAOT PES are showed in the table
\ref{coefgaot}. We plot in the figure \ref{sep1} the GAOT PES, considering $\theta$=30.0$^{\circ}$(a),
$\theta$=180.0$^{\circ}$(b) and  $\theta$=77.2$^{\circ}$ (c), with the respectives isoenergetic contours. The
dashed contours are taken between -160 and -40 kcal/mol and them are spaced each other by 5 kcal/mol. One can
see that the global shape of the GAOT PES is closed similar to both GSA and BO5 PES \cite{gargano5}.

\begin{center}
\begin{figure}[h]
\includegraphics{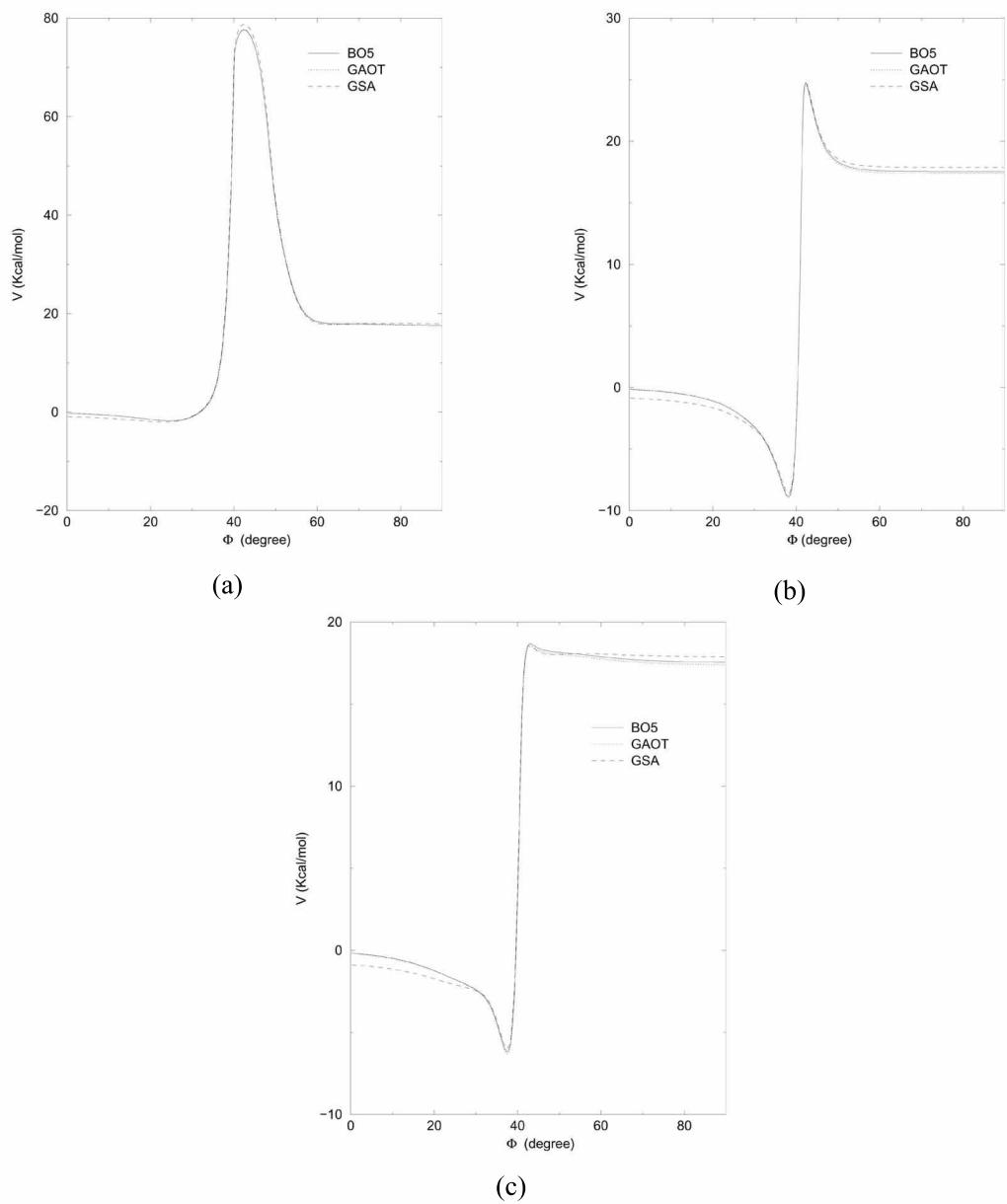}
\caption{The figure shows the plots of the BO5, GSA and GAOT  MEP for the Na+HF reaction calculated at
$\theta=$30.0$^{\circ}$(a), $\theta=$180.0$^{\circ}$(b) and $\theta=$77.2$^{\circ}$ (c). The $\Phi$ angle is
associated with the definition of the MEP (see text for discussion).} \label{mep1}
\end{figure}
\end{center}

To better test the new PES, we plot in figure \ref{mep1} the GAOT fixed angle minimum energy paths (GAOT MEP) of
the Na+HF reaction considering $\theta$=30.0$^{\circ}$ (a), $\theta$=180.0$^{\circ}$ (b) and
$\theta$=77.2$^{\circ}$ (c), respectively. In these figures are also shown both GSA and BO5 MEPs. In all these
MEPs the zero energy was set at the BO5 Na+HF asymptote. All these figures show that overall shape of the GAOT,
BO5 and GSA MEP are very similar. At $\theta$=77.20$^{\circ}$ the barrier of the BO5 reaction is minimum and
increases when moving towards collinear or towards more bent geometries. The same value was found for GAOT PES.
In the table \ref{meptable} are represented the values of the reactant energy, product energy, barrier height
and well depth of the BO5, GSA and GAOT MEPs considering the following values of the $\theta$=30$^{\circ}$,
60$^{\circ}$, 77.20$^{\circ}$, 90$^{\circ}$, 120$^{\circ}$, 150$^{\circ}$ and 180$^{\circ}$. In the region of
the Na+HF reactant, the differences of the energies find between the BO5-GAOT and GSA-GAOT MEP, for all values
of the $\theta$(see the \ref{meptable}), are about 0.02 kcal/mol and 0.73 kcal/mol, respectively. However, in
the NaF+H product region these differences are about 0.14 and 0.48 kcal/mol, respectively. In the well region,
the maximum differences of the energies between BO5-GAOT and GSA-GAOT are about 0.26 and 0.52 kcal/mol,
respectively. The maximum values of these differences in the barrier region are about 0.5 and 1.2 kcal/mol,
respectively.

\section{Conclusions}
\label{concl}

In this work, we have presented the GAOT method as a new option to fit PES for reactive system. Plots of the GAOT PES
and  BO5, GSA and GAOT MEP of the Na+HF system were made considering different values of the $\widehat{NaFH}$ angle.
From comparison among these plots, we concluded that these PES have the same global shape. The values of
the reactant energy, product energy, barrier height and well depth of these MEP at the principal
$\widehat{NaFH}$ angle considered were very small, within of the error acceptable to reactive process. This ample
topologies studies reveals that quality of the GAOT PES is comparable the with the best PES find for Na+HF reaction,
i.e, BO5 and GSA PES.

In a future work, we will present a complete study of the dynamics properties of the Na+HF reaction utilizing
the GAOT PES. These properties will be compared with the dynamics properties determined with the both BO5 GSA
PES.

\section{Acknowledgments}

This work has been supported by Brazilian Science and Technology Council (CNPq) and CAPES.

\begin{center}
\begin{table}[h]
\begin{tabular}{cccccccccc}
\hline \hline
\begin{tabular}{c}{\bf a$_{xyz}$}\\ \end{tabular} &
\begin{tabular}{c}{\bf x}\\ \end{tabular} &
\begin{tabular}{c}{\bf y}\\ \end{tabular} &
\begin{tabular}{c}{\bf z}\\  \end{tabular} &&
\begin{tabular}{c}{\bf a$_{xyz}$}\\ \end{tabular} &
\begin{tabular}{c}{\bf x}\\ \end{tabular} &
\begin{tabular}{c}{\bf y}\\ \end{tabular} &
\begin{tabular}{c}{\bf z}\\  \end{tabular} &
\\
\hline
-.306652213x10$^3$ & 1 & 0 & 0  &&  .154007168x10$^2$ &  1 &  2 &  2 \\
.340192379x10$^3$ & 2 & 0 & 0  && -.373620122x10$^1$ &  1 &  2 &  3 \\
-.255607454x10$^3$ & 3 & 0 & 0  && -.228197478x10$^2$ &  1 &  3 &  0 \\
.982458048x10$^2$ & 4 & 0 & 0  && -.747920955x10$^2$ &  1 &  3 &  1 \\
-.293524473x10$^3$ & 0 & 1 & 0  &&  .145451925x10$^2$ &  1 &  3 &  2 \\
.177532304x10$^3$ & 0 & 2 & 0  &&  .267614848x10$^3$ &  1 &  4 &  0 \\
-.39324702x10$^2$ & 0 & 3 & 0  &&  .163502219x10$^2$ &  1 &  4 &  1 \\
.141315171x10$^2$ & 0 & 4 & 0  && -.722678436x10$^2$ &  1 &  5 &  0 \\
-.989060904x10$^2$ & 0 & 0 & 1  && -.781884174x10$^3$ &  2 &  0 &  1 \\
.656454212x10$^2$ & 0 & 0 & 2  &&  .280647062x10$^3$ &  2 &  0 &  2 \\
-.14817526x10$^2$ & 0 & 0 & 3  &&  .571812073x10$^2$ &  2 &  0 &  3 \\
.302807924x10$^1$ & 0 & 0 & 4  && -.434796038x10$^2$ &  2 &  0 &  4 \\
.141262929x10$^2$ & 0 & 1 & 1  && -.103732088x10$^4$ &  2 &  1 &  0 \\
-.907353392x10$^1$ & 0 & 1 & 2  &&  .290375682x10$^3$ &  2 &  1 &  1 \\
.618984796x10$^2$ & 0 & 1 & 3  && -.964447908x10$^2$ &  2 &  1 &  2 \\
-.118497362x10$^2$ & 0 & 1 & 4  && -.743005471x10$^1$ &  2 &  1 &  3 \\
-.707766194x10$^1$ & 0 & 1 & 5  &&  .105653468x10$^4$ &  2 &  2 &  0 \\
.692426579x10$^2$ & 0 & 2 & 1  && -.391133693x10$^2$ &  2 &  2 &  1 \\
-.125374613x10$^3$ & 0 & 2 & 2  && -.796425832x10$^1$ &  2 &  2 &  2 \\
-.152740958x10$^2$ & 0 & 2 & 3  && -.416672911x10$^3$ &  2 &  3 &  0 \\
.86466874x10$^1$ & 0 & 2 & 4  && -.222050269x10$^2$ &  2 &  3 &  1 \\
.607286269x10$^2$ & 0 & 3 & 1  &&  .580961193x10$^2$ &  2 &  4 &  0 \\
.773420092x10$^2$ & 0 & 3 & 2  &&  .612684773x10$^3$ &  3 &  0 &  1 \\
-.100454752x10$^2$ & 0 & 3 & 3  && -.245706225x10$^3$ &  3 &  0 &  2 \\
-.844822048x10$^2$ & 0 & 4 & 1  &&  .537198805x10$^2$ &  3 &  0 &  3 \\
-.107302707x10$^2$ & 0 & 4 & 2  &&  .520636389x10$^3$ &  3 &  1 &  0 \\
.199310148x10$^2$ & 0 & 5 & 1  && -.11485817x10$^3$ &  3 &  1 &  1 \\
.43595539x10$^3$ & 1 & 0 & 1  &&  .378941105x10$^0$ &  3 &  1 &  2 \\
-.153727626x10$^3$ & 1 & 0 & 2  && -.312500339x10$^3$ &  3 &  2 &  0 \\
-.234081744x10$^2$ & 1 & 0 & 3  &&  .293249874x10$^2$ &  3 &  2 &  1 \\
-.493407059x10$^1$ & 1 & 0 & 4  &&  .633680268x10$^2$ &  3 &  3 &  0 \\
.992614849x10$^1$ & 1 & 0 & 5  && -.187852456x10$^3$ &  4 &  0 &  1 \\
.983517263x10$^3$ & 1 & 1 & 0  &&  .171140643x10$^2$ &  4 &  0 &  2 \\
-.346826233x10$^3$ & 1 & 1 & 1  && -.15087795x10$^3$ &  4 &  1 &  0 \\
.972616646x10$^2$ & 1 & 1 & 2  &&  .479562662x10$^2$ &  4 &  1 &  1 \\
-.250389965x10$^2$ & 1 & 1 & 3  &&  .23152828x10$^2$ &  4 &  2 &  0 \\
.246853766x10$^2$ & 1 & 1 & 4  &&  .27726868x10$^2$ &  5 &  0 &  1 \\
-.861838273x10$^3$ & 1 & 2 & 0  &&  .126767492x10$^2$ &  5 &  1 &  0 \\
.171256749x10$^3$ & 1 & 2 & 1 \\
\hline \hline
\end{tabular}
\caption{} \label{coefgaot}
\end{table}
\end{center}

\begin{table}[h]
\begin{tabular}{ccccccc}
\hline \hline
\begin{tabular}{c}{\bf $\Theta$}\\ \end{tabular} &
\begin{tabular}{c}{\bf MEP}\\ \end{tabular} &
\begin{tabular}{c}{\bf Reactant}\\ \end{tabular} &
\begin{tabular}{c}{\bf Product}\\  \end{tabular} &
\begin{tabular}{c}{\bf Barrier}\\  \end{tabular} &
\begin{tabular}{c}{\bf Well}\\  \end{tabular} &
\\
\hline
& BO5   & -0.2303 & 17.6297 & 77.6465 & -1.7699 \\
{\Large $_{30}$$^{\circ}$} & GSA  & -0.9266 & 17.9334 & 78.7675 & -2.0017 \\
 & GAOT & -0.2144 & 17.4871 & 77.5946 & -1.7776 \\
\\
 & BO5   & -0.1978 & 17.5697 & 23.2639 & -3.1729 \\
 {\Large $_{60}$$^{\circ}$}& GSA  & -0.9001 & 17.8983 & 23.3119 & -3.0291 \\
 & GAOT & -0.1819 & 17.4256 & 23.0681 & -3.3211 \\
\\
 & BO5   & -0.1817 & 17.5518 & 18.6754 & -6.1842 \\
 {\Large $_{77.20}$$^{\circ}$}& GSA  & -0.8869 & 17.8874 & 18.5825 & -5.9927 \\
 & GAOT & -0.1659 & 17.4072 & 18.5229 & -6.3219 \\
\\
 & BO5   & -0.1724 & 17.5444 & 18.9925 & -7.3662 \\
 {\Large $_{90}$$^{\circ}$}& GSA  & -0.8793 & 17.8829 & 19.4903 & -7.1637 \\
 & GAOT & -0.1565 & 17.3996 & 19.4439 & -7.4910 \\
\\
 & BO5   & -0.1586 & 17.5371 & 22.7954 & -8.3329 \\
 {\Large $_{120}$$^{\circ}$}& GSA  & -0.8681 & 17.8784 & 22.7551 & -8.1231 \\
 & GAOT & -0.1427 & 17.3996 & 22.6442 & -8.4518 \\
\\
 & BO5   & -0.1525 & 17.5351 & 24.3159 & -8.7335 \\
 {\Large $_{150}$$^{\circ}$}& GSA  & -0.8631 & 17.8772 & 24.3115 & -8.4731 \\
 & GAOT & -0.1366 & 17.3889 & 24.5618 & -8.9936 \\
\\
 & BO5   & -0.1508 & 17.5347 & 24.7239 & -8.8681 \\
 {\Large $_{180}$$^{\circ}$}& GSA  & -0.8617 & 17.8769 & 24.7062 & -8.6105 \\
 & GAOT & -0.1349 & 17.3884 & 24.5618 & -8.9936  \\
\hline \hline
\end{tabular}
\caption{} \label{meptable}
\end{table}

\end{document}